\documentclass[aps,pra,showpacs,twocolumn,superscriptaddress]{revtex4} 
\usepackage{graphicx,color}
\usepackage{amsmath,amsthm,amsfonts,amssymb,bm}
\usepackage{times}
\usepackage{epsf}
\usepackage[colorlinks={true}]{hyperref}
\usepackage{mathrsfs}
\usepackage[T1]{fontenc}
\usepackage[utf8]{inputenc}
\usepackage{xcolor}
\usepackage{appendix}

\hypersetup{citecolor={blue}, filecolor={blue}, linkcolor={blue}, urlcolor={blue}}

\newcommand{\commentold}[1]{}
\DeclareMathSymbol{:}{\mathpunct}{operators}{"3A}
\begin{document}
\title{Bounds on charging power of open quantum batteries}
\author{Shadab Zakavati}
\email{Sh.zakavati@uok.ac.ir}
\affiliation{Department of Physics, University of Kurdistan, P.O.Box 66177-15175, Sanandaj, Iran}
\author{Fatemeh T. Tabesh}
\affiliation{Department of Physics, University of Kurdistan, P.O.Box 66177-15175, Sanandaj, Iran}
\author{ Shahriar Salimi}
\email{ShSalimi@uok.ac.ir}
\affiliation{Department of Physics, University of Kurdistan, P.O.Box 66177-15175, Sanandaj, Iran}
\date{\today}
\begin{abstract} 
In general, quantum systems most likely undergo open system dynamics due to their smallness and sensitivity. Energy storage devices, so-called quantum batteries, are not excluded from this phenomenon. Here, we study fundamental bounds on the power of open quantum batteries from the geometric point of view. By defining an \emph{activity operator}, a tight upper bound on the charging power is derived for the open quantum batteries in terms of the fluctuations of the activity operator and the quantum Fisher information. The variance of the activity operator may be interpreted as a generalized thermodynamic force, while the quantum Fisher information describes the speed of evolution in the state space of the battery. The thermodynamic interpretation of the upper bound is discussed in detail. As an example, a model for the battery, taking into account the environmental effects, is proposed, and the effect of dissipation and decoherence during the charging process on both the stored work and the charging power is investigated. Our results show that the upper bound is saturated in some time intervals. Also, the maximum value of both the stored work and the corresponding power is achieved in the non-Markovian underdamped regime.
\end{abstract}
\pacs{03.65.Yz, 42.50.Lc, 03.65.Ud, 05.30.Rt}
\maketitle
\section{Introduction}
Recently, there has been a great deal of interest in studying quantum thermodynamics by increasing requests for device miniaturization \cite{thermo1, thermo2, thermo3}. The study of thermodynamic concepts in a quantum context is of great importance, both from a fundamental and a practical point of view \cite{thermo4, thermo5, thermo6, thermo7}. One of the main purposes of these blossoming researches is to propose various mechanisms and design devices to store and transfer energy beyond the microscopic scale, functioning as a battery. Accordingly, Quantum batteries ($QB$s) are introduced as finite dimensional quantum devices that are able to temporarily store energy in quantum degrees of freedom and transfer the energy to other apparatus~\cite{b1, b2, b3, b4, b5, b6,b7, b8, b9}. $QB$s have yet been suggested in a number of models, such as spin systems \cite{spin}, quantum cavities \cite{b2, b4, cavity}, superconducting transmon qubits \cite{transman}, Josephson quantum phase battery \cite{phase}, molecular battery \cite{mole}, Sachdev-Ye-Kitaev model \cite{sh}, and quantum oscillators \cite{ b7, harmo}.  

In most literature, $QB$s are regarded as closed systems, which follow entirely the unitary evolution. However, concerning the fragile nature of all quantum systems, it sounds plausible that batteries may interact with the surrounding environment, leading to the dissipation of the stored energy. To deal with this issue, the concept of open quantum batteries ($OQB$s) has been introduced in recent years \cite{transman, A2, A3, A4, bar, measur, dark, kamin,op, op1}. The evolution of $OQB$s can be characterized by means of a family of completely positive and trace-preserving maps. Consequently, $OQB$s dynamics can be either Markovian or non-Markovian \cite{n1,n2,n3}. The interaction of the $OQB$s with their reservoirs can lead to energy dissipation and decoherence. Hence, it is essential to find strategies to stabilize the energy storage against energy leakage into an environment \cite{transman, measur, dark, kamin}. To suppress these unwanted effects, recent research efforts have been devoted to retaining energy with minimized dissipation in $OQB$s \cite{transman, A2, A3, A4, bar, measur, dark, kamin}. 

In the study of the thermodynamic behavior of quantum systems in the context of quantum thermodynamics, the key argument is to derive a consistent formulation for the desired thermodynamic quantities from the acknowledged quantum principles. One of the main issues in the thermodynamic characterization of $QB$s is energy deposition referred to as charging process during which the state of a system is transferred from lower to higher energy levels. In general, the $QB$ charging protocol is composed of a $QB$ and a charger (energy source), where energy flows from the charger into the battery by establishing an interaction between them. Minimizing the charging time and maximizing the associated power are figures of merit. The Heisenberg's uncertainty relation as an essential and broadly-used principle in quantum theory, in a practical statement known as quantum speed limit ($QSL$), is interpreted as setting a fundamental bound on the intrinsic time scale of any quantum evolution. In other words, time-energy uncertainty quantifies how fast a quantum system can evolve \cite{def-camb}. $QB$ Hamiltonian quantifies the amount of energy that can be deposited in the battery. Due to the fact that battery Hamiltonian has finite magnitude, it holds a fundamental bound on the minimum time required to transform a given initial state to a given final state. In the framework of $QSL$ and geometry of quantum space, the minimized evolution time is obtained when the dynamical trajectory reaches to the \emph{geodesic} path. The geodesic path denotes the shortest length among all physical evolution trajectories between the given initial and final states. As the space of quantum states equipped with the proper metric, the geodesic distance can be calculated. Here, we will be working with the Bures metric in which the corresponding geodesic distance is known\cite{pires}.  

So far, a few different bounds have been introduced for the charging power. By means of a quantum geometrical approach, Farré et al.  \cite{bound} proposed a bound in terms of the energy variance of the battery and the Fisher information in the eigenspace of the battery Hamiltonian for closed $QB$s. In another study conducted by Pintos et al.\cite{op1}, the bound was defined in terms of the interaction Hamiltonian fluctuations and \emph{free energy operator} fluctuations. The later was shown to be valid for closed $QB$s as well as $OQB$s. Furthermore, they concluded that there must exist fluctuations in the extractable work stored in the battery to have a non-zero charging power. Also, in an interesting work, exploiting the notion of $QSL$, an upper bound for the charging power of arrays of $N$ batteries was proposed based on the quantum collective effects \cite{b3}.

Motivated by the above considerations and recent progress in $OQB$s, this study aims to answer the following questions: Is it possible to generalize a bound on the charging power for $OQB$s in the context of the geometry of quantum states? if so, what is the thermodynamic description of terms appear in this bound? Can one engineer a dissipative charging process for a battery and keep the stored energy stabilized by using quantum memory effects? To address the questions, we study bounds on the charging power and generalize the previous bounds. Due to the fact that every system out of equilibrium in contact with a thermal bath contains an amount of free energy that can do work, we define an \emph{activity operator} which quantifies how far the state of the system distances from equilibrium. A tight upper bound on charging power in terms of quantum Fisher information ($QFI$) and the variance of the activity operator of $OQB$s is proved. By dividing the dissipation part of Lindblad master equation into a diagonal part and a non-diagonal part, a redefinition for the bound in terms of dissipative work and entropy production rate is proposed. By applying the notion of the extended quantum Fisher information, the speed of evolution is divided into the classical and quantum parts, based on more physically meaningful contributions.  Moreover, the role of dissipation effects and the backflow of information on the stored work and charging power is explored. For this, an example will be considered in which a battery interacts with dissipative and heating reservoirs at finite temperature. We will show that the stored work and the power are maximal for non-Markovian underdamped regime. Results indicate that the charging power of $OQB$ can boost by increasing temperature if interaction parameters, coupling coefficient, and temperature are adjusted properly. despite the fact that one may anticipate the performance of $QB$s can be spoiled at high temperatures.

The paper is organized as follows. In Sec. II bounds on the charging power of open systems are provided. The derivations in Sec. III. sheds light on the Thermodynamics interpretation of the bound. In Sec. IV, an upper bound on the charging power is suggested based on an extended QFI. In order to illustrate the upper bound is tight, Heisenberg XX spin chain example is presented in Sec. V. Also, a heuristic model of OQB is investigated in the presence of a bath. The conclusion is summarized in Sec. VI.

\section{Bounds on charging power}

First, a general model describing $OQB$s is presented. The model is constructed from a quantum system as a battery and a charging protocol. The battery system also interacts with a thermal bath in the framework of open system analysis. The Hamiltonian of the whole system of the charger $A$ , the battery $B$ and the bath $E$ is defined by
\begin{eqnarray}\label{e11}
H= H_{A}+ H_{B}+ H_{E}+H_{int},
\end{eqnarray}
where $H_{A}$,  $H_{B}$ and  $H_{E}$ are the charger, the battery and the bath free Hamiltonians, respectively.
$H_{int}$ includes all interactions with the $QB$.  Note that, $H_{B}$ is time independent. Therefore, in the interaction picture, the reduced density matrix of the $QB$ at time $t$ can be written as
\begin{eqnarray}\label{e12}
\partial_{t}\rho= -i~ Tr_{(AE)}[H_{int},\rho_{tot}],
\end{eqnarray}
in which, the density matrix $\rho_{tot}$ on the right side refers to the total state, including battery system+charger+environment and the partial trace is taken over the charger and the environment subsystems. For a system in contact with a thermal bath, every state of the system out of equilibrium contains an amount of free energy that can be extracted in the form of work. The non-equilibrium free energy is defined as
\begin{eqnarray}\label{e2}
F(\rho)=U-\beta^{-1}S(\rho),
\end{eqnarray}
in which, $U=Tr(\rho H_{B})$ and $S(\rho)=-Tr(\rho \ln\rho)$ are respectively the energy and von Neumann entropy of the system, and $\beta=T^{-1}$ denotes the inverse temperature of bath \cite{alle, P, FG}. Here, Boltzmann's constant $k_{B}$ is set equal to 1, $k_{B}=1$ as a convention. In the relaxation process, the free energy of the system naturally tends to decrease until it reaches its minimum value. The equilibrium state, therefore, denotes the state at which the free energy is minimized. By assuming that the instantaneous state of the $QB$ is $\rho$, and the thermal equilibrium state is indicated by $\tau_{\beta}$. The maximum extractable work from the battery system is given by
\begin{equation}\label{e1}
W_{max}= F(\rho)-F(\tau_{\beta}),
\end{equation}
where $\tau_{\beta}=\frac{1}{Z} \exp (-\beta H_{B})$ and $ Z= Tr(\exp (-\beta~H_{B})$ is the partition function. In the following, by defining $\mathscr{A}:=\beta^{-1}log \left ( \dfrac{\rho}{\tau_{\beta}}\right )$ as  the activity operator, we can rewrite the maximum extractable work as
\begin{equation}\label{e3} 
W_{max}= \frac{1}{\beta}Tr(\rho (\ln \rho-\ln\tau_{\beta} ))=Tr(\rho ~\mathscr{A}).
\end{equation}
When the system is at equilibrium, obviously, $\mathscr{A}=0$ and $\mathscr{A}>0$ for any other non-equilibrium state. The activity operator quantifies how far the state of the system distances from equilibrium. In other words, activity operator associated with a state of the system indicates how much the state is \emph{active} or has an availability to extract work from it. It is worth noting that in \cite{op1}, $\mathbb{F}= H_{B}+\beta^{-1} \log \rho$  has been introduced as \emph{work operator}.

How fast the work can stored on the $QB$ depends on its charging power, i.e., the rate at which the energy flows in the $QB$ during the interaction. The charging power is determined by

\begin{equation}\label{e5}
\frac{d}{dt}(W_{max})= \frac{d}{dt}Tr(\rho ~\mathscr{A}) = Tr(\dot{\rho}~\mathscr{A})+Tr(\rho~\dot{\mathscr{A}}),
\end{equation}
by noting that the time dependency of activity operator $\mathscr{A}$ is exclusively due to state $\rho $ of battery. The external time-dependent agent acts on a finite time, and the thermal states do not take into account such an external field. So, the free Hamiltonian of the battery, which determines the structure of the battery, acts as a thermalization Hamiltonian. As a result, the second derivative at the left side of the equation of no time dependence is considered in the Gibbs thermal state.
 as a result,
\begin{eqnarray}\label{e6}
\mathbb{P}= Tr(\dot{\rho}~\mathscr{A}).
\end{eqnarray}
Now, based on the above formula, an upper bound for charging power of a $OQB$ is found. As mentioned earlier, the upper bound on the power saturates when the time required to transform a given initial state to a given final state is minimized. It occurs when among all the possible dynamical trajectories, the system evolves through the geodesic path, which is the shortest curve between two distinguishable states.  The distinguishability of quantum states can be characterized by a distance measure on density operator space.
The distance between two neighbouring points on the manifold of quantum states induces a metric $g_{tt}$, that can be written as $D^{2}( \rho(t+dt)-\rho(t))=g_{tt}dt^{2}$, where $D$ is a given distance in the state space \cite{vall}. From this, one can simply conclude that $\sqrt{g_{tt}}$ can be interpreted as the instantaneous speed of evolution of $\rho_{t}$. In the present study, we consider Bures distance, which has the advantage that whose geodesic is analytically known and equivalent to quantum fisher information  metric.
A well-known statement for the $QFI$ can be provided by the use of symmetric logarithmic derivative
\begin{equation}\label{e7}
I_{Q}(\diamond)=Tr(\rho ~ L(\diamond)^{2}),
\end{equation}
in which, $\diamond$ denotes the desired parameter and the Hermitian operator $L$ for a given state $\rho(t)$ and $t$ as a parameter is defined through \cite{fisher}
\begin{equation}\label{e8}
\partial_{t}\rho=\frac{1}{2}( L~\rho+\rho ~ L).
\end{equation}
Note that the parameter dependence is omitted to simplify the notation. Now, we can get back to Eq. (\ref{e6}) and rewrite it as

\begin{equation}\label{e8-1}
|\mathbb{P}|=|\operatorname{Tr}(\dot{\rho} \mathscr{A})|=|\operatorname{Tr}(\dot{\rho}(\mathscr{A}-\langle\mathscr{A}\rangle))|=|\operatorname{Tr}(\dot{\rho} \delta\mathscr{A})|.
\end{equation}
Note that $\operatorname{Tr}(\dot{\rho}\langle\mathscr{A}\rangle)=\langle\mathscr{A}\rangle \operatorname{Tr}(\dot{\rho})=0$ and also by definition $\delta \mathscr{A}=\mathscr{A}-\langle\mathscr{A}\rangle$. By replacing  Eq. (\ref{e7}) and Eq. (\ref{e8}) in the above formula, the power can be written as

\begin{equation}\begin{aligned}\label{e9}
&\vert\mathbb{P}\vert= \vert Tr(\frac{1}{2}( L~\rho+\rho ~L)~\delta\mathscr{A})\vert\\
&\leq \frac{1}{2}\vert Tr( L~\rho ~ \delta \mathscr{A})\vert+ \frac{1}{2}\vert Tr(\rho ~ L ~ \delta \mathscr{A})\\
&~=\operatorname{Tr}(L \rho \delta \mathscr{A}).
\end{aligned}\end{equation}
The second line follows from triangle inequality. The third line is due to the fact that for any Hermitian operator A,B and C,  $|\operatorname{Tr}(A B C)|=|\operatorname{Tr}(A C B)|$. Cyclic property of trace implies that $|\operatorname{Tr}( L~\rho ~ \delta \mathscr{A})|=\operatorname{Tr}( \sqrt{\rho}L~ \delta \mathscr{A}~\sqrt{\rho})|$, by noting that for the positive operator of $\rho$ there exists a square root operator $\sqrt{\rho}$. Finally, by using the Cauchy-Schwarz inequality $\left|\operatorname{Tr}\left(A^{\dagger} B\right)\right|^{2} \leq \operatorname{Tr}\left(A^{\dagger} A\right) \operatorname{Tr}\left(B^{\dagger} B\right)$, we can obtain the following inequality

\begin{eqnarray}\label{e10}
|\mathbb{P}| \leq \sqrt{\operatorname{Tr}\left(\rho L^{2}\right) \operatorname{Tr}\left(\rho \delta \mathscr{A}^{2}\right)}=\sigma_{\mathscr{A}} ~\sqrt{I_{Q}},
\end{eqnarray}
in which $\sigma_{\mathscr{A}}$ indicates standard deviations of activity operator. The above inequality shows an upper bound on the charging power, which generalizes the bound proposed for closed QBs~\cite{bound} to $OQBs$. As mentioned earlier, the square root of the quantum Fisher information represents the speed of evolution in the state space of the battery. Therefore, an immediate insight from Eq. (\ref{e10}) reminds us of the familiar formula of power in classical physics $P\propto F.v$, where $F$ can be any (constant) force and $v$ is the flow velocity relative to the object. therefore, In comparison with this formula, the variance of the activity operator may be characterized as generalized thermodynamic force. The activity operator associated with a non-equilibrium state, therefore, drives the system towards the equilibrium state. A similar statement for the thermodynamic force provided in~\cite{borhan1}. It is expected that at equilibrium, $\mathscr{A}$ as all thermodynamic forces must vanish. An example to further clarify this phenomenon is the temperature gradient which can be regarded as a thermodynamic force that causes an irreversible flow of heat between two systems until they reach the same temperature. 

In the next section by using the Lindblad type master equation, we obtain thermodynamic interpretation of the upper bound in terms of  the dissipative work and the entropy production rate.

\section{Thermodynamic interpretation of the bound}\label{sec2}
Having introduced the model of $OQB$ and a definition for power, we explain the bound in terms of thermodynamic arguments. To this aim, by taking the partial trace over the bath and charger in Eq. (\ref{e12}), the reduced  dynamics of the $QB$ can be described by the following
 master equation \cite{n1}
\begin{eqnarray}\label{e17}
\partial_{t}\rho=-i~[H_{B}(t),\rho(t)]+D[\rho(t)],
\end{eqnarray}

where the first term represents the unitary part of the dynamics. The term $ D[\rho(t)] $ represents the quantum dissipator which is defined as 
\begin{eqnarray}\label{e18}
D[\rho(t)] =\sum_{\alpha} \gamma_{\alpha}(t)[L_{\alpha}\rho(t)L^{\dag}_{\alpha}-\frac{1}{2}\{L^{\dag}_{\alpha}L_{\alpha}, \rho(t)\}] ,
\end{eqnarray}
 in which $ L_{\alpha} $ and $\gamma_{\alpha}(t) $  are Lindblad operators and decay rates, respectively.
 
In the following, by considering the spectral decomposition of the density matrix, i.e. $\rho(t)=\sum_{n}P_{n}(t) \vert n(t)\rangle \langle n(t)\vert$, the dissipator can be split as \cite{funo}
\begin{eqnarray}\label{e19}
D[\rho(t)] =D_{d}[\rho]+D_{nd}[\rho],
\end{eqnarray}
where the diagonal part is
\begin{eqnarray}\label{e20}
&&D_{d}[\rho]= \sum_{n} \langle n|D[\rho]| n\rangle |n\rangle\langle n|\nonumber \\
&&=\sum_{n}\partial_{t}P_{n}|n\rangle\langle n|=- ~\{ \Gamma(t),\rho(t) \},
\end{eqnarray}
in which, $\Gamma(t)$is defined as
\begin{eqnarray}\label{e21} 
\Gamma(t)=- \frac{1}{2}\sum_{n}\frac{\partial_{t}P_{n}}{P_{n}}|n\rangle\langle n|,
\end{eqnarray}
as suggested in \cite{alipo}. The non-diagonal part of the dissipator can be written as 
\begin{eqnarray}\label{e22}
D_{nd}[\rho]= \sum_{n\neq m} \langle n|D[\rho(t)]| m\rangle |n\rangle\langle m|.
\end{eqnarray}
By introducing the dissipative Hamiltonian 
\begin{eqnarray}\label{e23}
H_{Diss}(t)= \sum_{n\neq m} \frac{i ~\langle n|D[\rho]| m\rangle }{P_{m}-P_{n}}|n\rangle\langle m|,
\end{eqnarray}
thus, Eq. (\ref{e22}) takes the following form
\begin{eqnarray}\label{e24}
D_{nd}[\rho]=- i~[H_{Diss}(t),\rho(t)].
\end{eqnarray}
The above equation can be expounded as part of the bath dynamics which generates a unitary time-evolution \cite{funo}.

As a result, the Lindblad master equation can be written as
 \begin{eqnarray}\label{e25}
\partial_{t}\rho=-i~[\tilde{H},\rho(t)]-\{ \Gamma(t),\rho(t) \},
\end{eqnarray} 
where $ \tilde{H}=H(t)+H_{Diss}(t) $.

In the following, substituting the Eq. (\ref{e25}) into Eq. (\ref{e6}), one can find  
\begin{eqnarray}\label{e26}
\vert\mathbb{P}\vert = \vert Tr(-i ~[\tilde{H}, \rho]\mathscr{A})+Tr(-~\lbrace \Gamma(t), \rho\rbrace\mathscr{A}) \vert\nonumber \\
~~\leq \vert Tr(-i ~ [\tilde{H}, \rho]\mathscr{A}) \vert+\vert Tr (-~\lbrace \Gamma(t), \rho\rbrace\mathscr{A})\vert.
\end{eqnarray} 
Rewriting the activity operator as $\mathscr{A}= H_{B}+\beta^{-1} \log \rho+F_{\tau_{\beta}}$ (see Appendix \ref{appendix A}), one can obtain
\begin{eqnarray}\label{e27}
&&\vert\mathbb{P}\vert\leq\left\vert Tr([\rho,\tilde{H}]H_{B}) \right\vert \nonumber \\
&&+\left\vert Tr \left( \sum_{n}\dot{P_{n}}|n\rangle\langle n|\left( H_{B}+\beta^{-1} \log \rho \right) \right) \right\vert.\nonumber \\
\end{eqnarray}
The first term on the right-hand side of the above equation can be regarded as dissipative work
\begin{eqnarray}\label{e28}
&&\left\vert Tr([\rho,\tilde{H}]H_{B}) \right\vert= \left\vert Tr([\rho,H_{B}+H_{Diss}]H_{B})\right\vert \nonumber \\
&&=\left\vert Tr([\rho,H_{Diss}]H_{B})\right\vert=W_{Diss},
\end{eqnarray}
hence the commutator $[\rho,H_{Diss}]$ represents the unitary part of the dissipator. The second term of Eq. (\ref{e27})  can be written as
%
%\begin{eqnarray}\label{e29}
%&& \vert Tr \left( \sum_{n}\dot{P_{n}}|n\rangle\langle n|\left( \beta^{-1} \log\rho\right) \right) \nonumber \\ + Tr \left( \sum_{n}\dot{P_{n}}|n\rangle\langle n| H_{B}\right) \vert 
%&& =\left\vert -\frac{1}{\beta} dS+dQ\right \vert=\frac{1}{\beta} \vert dS_{irr}\vert.
%\end{eqnarray}
%
%
\begin{gather}
\bigg\vert Tr \left( \sum_{n}\dot{P_{n}}|n\rangle\langle n|\left( \beta^{-1} \log\rho\right) \right) \notag \\
+ Tr \left( \sum_{n}\dot{P_{n}}|n\rangle\langle n| H_{B}\right) \bigg\vert  
=\left\vert -\frac{1}{\beta} dS+dQ\right \vert=\frac{1}{\beta} \vert dS_{irr}\vert.
\label{e29}
\end{gather}
presenting the change in the irreversible entropy $ S_{irr} $, \cite{borhan, reza}. Therefore, combining the equations (\ref{e27}), (\ref{e28}) and (\ref{e29}), the charging power is bounded from above  as the following form
\begin{eqnarray}\label{e30}
&&\vert\mathbb{P}\vert\leq W_{Diss}+\frac{1}{\beta}\vert dS_{irr}\vert.
\end{eqnarray}
In the next section, the upper bound in Eq. (\ref{e10}) will be illustrated by means of a Heisenberg XX
spin chain for three qubits, and quantum battery in dissipation/heating reservoir. We will see that the bound in Eq. (\ref{e10}) is saturated with these cases. In addition, we will study the role of non-Markovian effects on energy conservation and enhance of charging power. 
\begin{figure}[t]
\includegraphics[scale=0.45]{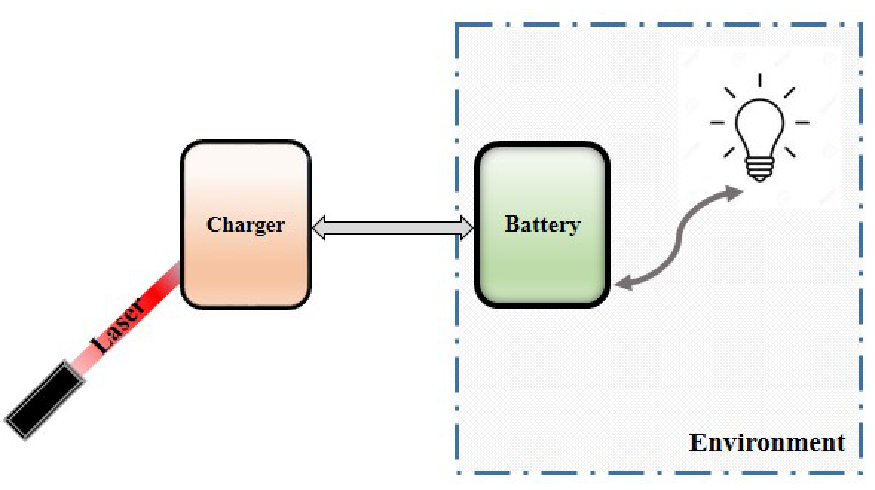}
\caption{Schematic diagrams of an open quantum battery. A quantum battery interacts with a charger during the charging process, while the battery is coupled individually into an environment. In addition, an external field is applied to the charger.}
\label{fig2}
\end{figure}

\section{Derivation of the Upper bound: extended QFI}

In Sec. II, the standard QFI was employed to derive an upper bound on charging power. Furthermore, splitting the dissipator into dissipative (non-unitary) and coherent (unitary) contributions allows one to write the whole master equation as commutator and anti-commutator parts, which is valid for the Lindblad-like master equations. By considering the time as a  parameter, such a decomposition of quantum Liouvillian has been based to introduce an extended QFI in terms of the non-Hermitian SLD \cite{nsld,alipo}. The extended QFI is defined as an upper bound on the QFI.

\begin{eqnarray}\label{301}
{I_{Q}}(x) \leq \operatorname{Tr}\left[\tilde{L}(x) \rho \tilde{L}(x)^{\dagger}\right],
\end{eqnarray}
in which, the nSLD $\tilde{L}$ satisfies $\partial_{x} \rho=\left(\tilde{L}(x) \rho+\rho \tilde{L}(x)^{\dagger}\right) / 2$.   
The right-hand side of the above inequality denoted by $I_{Q}^{e x t}$ Considering Eq. (\ref{e25}) and $t$ as the parameter, nSLD reads 
\begin{equation}\label{302}
\tilde{L}=-2 i(\tilde{H}-i \Gamma(t)).
\end{equation}
 In the following, by the similar procedure as in sec. II, a bound on charging power in terms of the extended QFI is obtained 

\begin{equation}\label{303}
P(t) \leq|T r[\dot{\rho} \mathscr{A}]|=| \operatorname{Tr}\left[\frac{\left(\tilde{L} \rho+\rho \tilde{L}^{\dagger}\right)}{2}\right] \delta \mathscr{A}|.
\end{equation}
The triangle and the Cauchy-Schwarz inequalities imply that

\begin{equation}\begin{aligned}\label{304}
&P(t) \leq \frac{1}{2}|\operatorname{Tr}(\operatorname{\tilde{L}\rho\delta} \mathscr{A})|+\frac{1}{2}\left|\operatorname{Tr}\left(\rho \tilde{L}^{\dagger} \delta \mathscr{A}\right)\right|\\
&=\frac{1}{2}|\operatorname{Tr}(\sqrt{\rho} \delta \mathscr{A} \tilde{L} \sqrt{\rho})|+\frac{1}{2}\left|\operatorname{Tr}\left(\sqrt{\rho} \tilde{L}^{\dagger} \delta \mathscr{A} \sqrt{\rho}\right)\right|\\
&\leq 2 \times \frac{1}{2} \sqrt{\operatorname{Tr}(\tilde{L} \rho \tilde{L}^{\dagger} \operatorname{Tr}\left(\rho \delta \mathscr{A}^{2}\right)}=\sqrt{{I_{Q}^{e x t}}} \sigma_{\mathscr{A}}.
\end{aligned}\end{equation}

In the above equation, the square root of $I_{Q}^{e x t}$ can be interpreted as the speed of evolution. In the following we are interested to separate this velocity term into contributions with certain physical interpretation. Substituting Eq.(\ref{302}) (in the interaction picture) into $I_{Q}^{e x t}$ gives
\begin{equation}\begin{aligned}\label{305}
&I_{Q}^{e x t}=4\operatorname{Tr}\left((H_{Diss}-i \Gamma(t)) \rho ({H_{Diss}}+i \Gamma(t))\right)\\
&=\operatorname{Tr}\left(\rho\left(H_{Diss}^{2}+\sum_{n}\left[\frac{\partial_{t} P_{n}}{P_{n}}\right]^{2}|n\rangle\langle n|\right)\right),
\end{aligned}\end{equation}

where in the second line we also have used Eq. (\ref{e21}). Here, One can show that the $\operatorname{Tr}\left(\rho\left(H_{Diss}\right)\right)=0$, therefore, the first term in the second line implies the variance of the $H_{Diss}$. By using of the spectral decomposition of the density matrix, we can conclude

\begin{equation}\label{306}
I_{Q}^{e x t}=\sigma_{H_{D iss}}^{2}+\sum_{n}\left[\frac{\partial_{t} P_{n}^{2}}{P_{n}}\right].
\end{equation}
The second term clearly represents the classical Fisher information. Thus, the fluctuations of the dissipative work denotes the pure quantum part of the extended QFI. This sounds sensible since the Hamiltonian of $H_{Diss}$ results from off-diagonal(coherent) part of dissipator. Note that The square root of QFI can be understood as the velocity at which system is transmitted between initial and final state. Therefore
\begin{equation}\label{307}
\sqrt{I_{Q}^{e x t}}=v(t)=\sqrt{v_{C L}(t)^{2}+v_{Q}(t)^{2}}.
\end{equation}

 As a conclusion, Eq. (\ref{307}) separates the speed of evolution into a classical and a quantum contribution. Each part relates to a physically meaningful quantity. In other words, the individual role of  populations of the state and the coherences in driving the evolution is clarified.

\section{Examples}\label{sec3}
\subsection{The Heisenberg XX spin chain}

Having established a framework for the charging power of OQBs, we now study the behavior of charging power and the corresponding upper bound in the following two examples. Our intention is to take into account the effect of the environment and different dynamical regimes.  For this, we consider simple illustrative models to illuminate key features of our framework. Note that here we exclusively use the upper bound derived based on the standard QFI in Sec II. The reason is that, according to  Eq (\ref{301}), this bound is expected to be tighter compared to the one defined based on the extended QFI in Sec IV (see Eq. (\ref{304})).

As the first example, we consider a three-qubit Heisenberg XX spin chain, where a qubit is regarded as the system and the other qubits as the environment and charger. The free Hamiltonian is  
\begin{equation}\label{e31}
H_{0}=\frac{\omega_{0}}{2}(\sum_{n=1}^{3}\sigma^{z}_{n}+1_{n}),
\end{equation}
where $\omega_{0}$ is the transition frequency of each qubit and for the sake of convenience, 
 ground-state energy is assumed to be zero.
The interaction Hamiltonian characterizing the chain exposed to a uniform magnetic field is given by
\begin{equation}\label{e32}
V=\frac{J}{2}\sum_{n=1}^{3}(\sigma^{x}_{n}\sigma^{x}_{n+1}+\sigma^{y}_{n}\sigma^{y}_{n+1})+ B\sum_{n=1}^{3} \sigma^{z}_{n},
\end{equation}
where  $\sigma^{\alpha}_{n}$ represents the Pauli operator corresponding to each $\alpha$ $(\alpha=x,y,z)$, $J$ marks the exchange interaction constant, and  $B$ is the magnitude of a uniform magnetic field \cite{tabesh}.
Suppose the periodic boundary conditions, $\sigma^{x}_{1}=\sigma^{x}_{4}$ and $\sigma^{y}_{1}=\sigma^{y}_{4}$, and consider eigenvalues and eigenstates of the Hamiltonian,
if the normalized initial taken as
\begin{equation}\label{e33}
|\Psi(0)\rangle=\alpha|001\rangle+\beta|010\rangle+\gamma|100\rangle,
\end{equation}
its time evolution will be
\begin{equation}\label{e34}
|\Psi(t)\rangle=a(t)|001\rangle+b(t)|010\rangle+c(t)|100\rangle,
\end{equation}
where
\begin{equation}\label{e35}
\begin{split}
a(t)= \frac{1}{3}( e^{it(J+B)}(2\alpha-\beta-\gamma)+ K(t)),\\
b(t)=\frac{1}{3}( e^{it(J+B)}(2\beta-\alpha-\gamma)+ K(t)),\\
c(t)=\frac{1}{3}( e^{it(J+B)}(2\gamma-\alpha-\beta)+ K(t)),
\end{split}
\end{equation}
in which $K(t)=e^{-it(2J-B)} (\alpha+\beta+\gamma)$.

Here, if we consider $ \rho= \sum_{n} P_{n} |n \rangle\langle n |$, with $0< P_{n} \leq 1$ and $\sum_{n} P_{n}=1$, the quantum Fisher information for the parameter $t$ can be calculated as \cite{fish}
\begin{eqnarray}\label{e36}
I_{Q}&=&\sum_{n}\frac{(\partial_{t}P_{n})^{2}}{P_{n}}+\sum_{n} 4P_{n} \langle \partial_{t}n| \partial_{t}n\rangle \nonumber \\
&-&\sum_{n, m}\frac{8P_{n}P_{m} }{P_{n}+P_{m}}|\langle\partial_{t}n|\ m\rangle|^{2}.
\end{eqnarray}

\begin{figure}[t]
\includegraphics[scale=0.56]{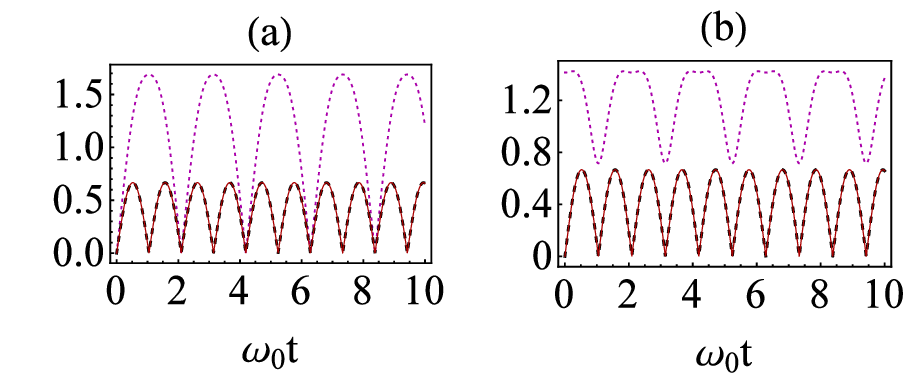}
\caption{(Color online). Plot of $(2 \sigma_{V}\sigma_{\mathbb{F}})/\omega_{0}$ (dotted magenta line) and  $(\sigma_{\mathscr{A}} \sqrt{I_{Q}})/\omega_{0}$ (dashed black line)  and $ |\mathbb{P}| /\omega_{0}$ (red solid line) as a function of $\omega_{0}t$. We have used  $\alpha=0$, $\beta=0$, and $\gamma=1$ in (a) and 
$\alpha=0$, $\beta=1/\sqrt{2}$, and $\gamma=1/\sqrt{2} $ in (b). Other parameters are $ J=\omega_{0}$, and $B=0 $.}
\label{fig1}
\end{figure}

Using Eqs. (\ref{e5}),  (\ref{e31}), (\ref{e34}) and the above equation, one can obtain the following equality
\begin{equation}\label{e37}
|\mathbb{P}|=\sqrt{I_{Q}}\sigma_{\mathscr{A}}=\omega_{0}\vert \dot{b}b^{\ast}+b \dot{b^{\ast}} \vert.
\end{equation}
From the equation above it can be seen that the equality holds over time which indicate that in this example the LHS of Eqs. (\ref{e10}) is saturated to the upper bound. For comparison with the bound introduced in Ref. \cite{op1}, the two upper bounds and $\vert\mathbb{P}\vert$ with respect to $\omega_{0}t$ are shown in Fig. \ref{fig1}, where dotted magenta line represents $(2 \sigma_{V}\sigma_{\mathbb{F}})/\omega_{0}$ , dashed black line shows $(\sigma_{\mathscr{A}} \sqrt{I_{Q}})/\omega_{0}$  and red solid lines indicate  $|\mathbb{P}|/\omega_{0} $.  We have $ \alpha=0,\beta=0,\gamma=1 $ in Fig. \ref{fig1}(a)
and $\alpha=0$, $\beta=1/\sqrt{2}$, $\gamma=1/\sqrt{2} $  in Fig. \ref{fig1}(b). As can be seen $\sigma_{\mathscr{A}} \sqrt{I_{Q}}$ is reached while $2 \sigma_{V}\sigma_{\mathbb{F}}$ is greater than
 $|\mathbb{P}|$. A proof, allowing the comparison between our bound and those suggested in Ref. \cite{op1}, is presented in Appendix \ref{appendix A}.
\subsection{Quantum battery and dissipative/heating reservoir}
As discussed before, the definition of the charging power and stored work are based on the activity operator, which depends on inverse temperature $\beta$ and the state $\rho$. The state of an open system at any time is the solution of the master equation in which the dissipation terms are included. Thus, the system's state seems to be sensitive to the environment and interaction parameters and conditions under which the master equation has been solved. Hence, before going through the details of following example and regardless of the complexity of the system under consideration, we expect that the effects of temperature and the environment parameters on the charging power and extractable work are important.
In the following, we assume a charging protocol where, the $QB$ is immersed in a reservoir including the effects of dissipation and heating (see Fig. \ref{fig2}).
Let us consider the case in which both the charger $A$ and the $QB$ are two qubits. The total Hamiltonian is expressed as \cite{op}
\begin{equation}\label{e38} 
H=H_{0}+\Delta H_{A}+H_{AB}+H_{BE},
\end{equation}
where the first term is  the free Hamiltonian of the total system given by
\begin{equation} \label{e39}
H_{0}=\frac{\omega_{0}}{2}(\sigma^{A}_{z}+1)+\frac{\omega_{0}}{2}(\sigma^{B}_{z}+1)+\sum_{k}\omega_{k}b^{\dagger}_{k}b_{k},
\end{equation}
and interaction Hamiltonians can be expressed as 
\begin{eqnarray} \label{e40}
\Delta H_{A}=\eta (\sigma^{A}_{+} e^{- i\omega_{0}t}+\sigma^{A}_{-}e^{ i\omega_{0}t}),\nonumber\\
H_{AB}=\kappa(\sigma^{A}_{+}\sigma^{B}_{-}+\sigma^{A}_{-}\sigma^{B}_{+}),\nonumber\\
H_{BE}=\sum_{k}g_{k}(\sigma^{B}_{+}b_{k} +\sigma^{B}_{-}b^{\dagger}_{k}).
\end{eqnarray} 
In the above equation, $\sigma^{A,B}_{\pm}$ is the raising and the lowering operators of the corresponding qubit,  $\omega_{0}$ and $\omega_{k}$ are respectively the transition frequency of the qubits and the environment; $ b_{k} $ ($b^{\dagger}_{k} $) represents the annihilation
(creation) operator corresponding to the $k$th mode of the bosonic environment; and
$ g_{k} $ indicates the coupling constant between the battery and  the $k$th mode of the environment.
The first term in Eq. (\ref{e40}), $\Delta H_{A}$, defines an external resonant driving field with amplitude $\eta$ that may inject energy into the system and the second term, $H_{AB}$, shows the interaction Hamiltonian between the charger and the battery by the coupling constant $ \kappa $. Finally, $ H_{BE}$ describes the interaction between the battery and the bath at temperature $T$. We emphasize that the charger does not couple to the bath.
\begin{figure}[t]
\includegraphics[scale=0.5]{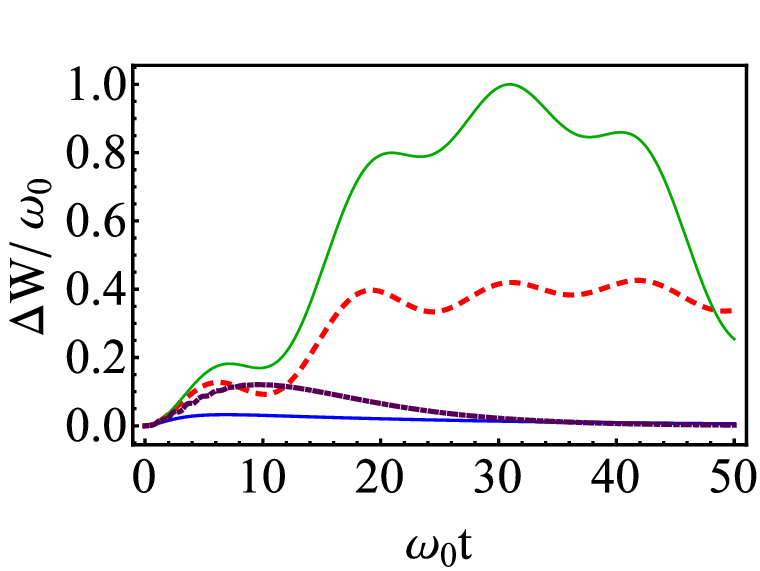}
\caption{(Color online).  Plot of $ \Delta W/ \omega_{0}$, (in units of $\omega_{0}$),  as a function of $\omega_{0}t$. Numerical results in this plot have been obtained by setting $ \eta=0.1\omega_{0} $, $ \Delta=3\omega_{0} $ and $ T=0 $.  Solid blue line
 (overdamped regime) and  dashed red line (underdamped regime) present local Markovian dynamics ($R=0.01$) for $ \gamma_{0}= \omega_{0}, \lambda=100 \omega_{0}$, $ \kappa=0.001\omega_{0} $ and $ \gamma_{0}= 0.1\omega_{0} , \lambda=10\omega_{0}$, $ \kappa=0.2\omega_{0} $, respectively. 
Dotted purple line (overdamped regime) and solid green  line (underdamped regime) remarks local non-Mmarkovian dynamics ($R=10$) for $ \gamma_{0}=10 \omega_{0} , \lambda=1 \omega_{0}$ , $\kappa=0.001\omega_{0} $ and $ \gamma_{0}= 0.1\omega_{0} , \lambda=0.01\omega_{0}$, $\kappa=0.2\omega_{0} $, respectively. 
We have considered $|\varphi_{AB}(0)\rangle= |1\rangle\otimes|0\rangle$. }
\label{fig3}
\end{figure}
 In the interaction picture representation, the corresponding master equation of the model explicitly reads as \cite{op, tabesh1}
\begin{eqnarray}\label{e41}
&&\frac{d\rho^{AB}}{dt}=-i[\kappa(\sigma^{A}_{+}\sigma^{B}_{-}+\sigma^{A}_{-}\sigma^{B}_{+})+\eta (\sigma^{A}_{+}+\sigma^{A}_{-}),\rho^{AB}]\nonumber\\
&&+\dfrac{\gamma_{1}(t)}{2}(\sigma^{B}_{+}\rho^{AB}\sigma^{B}_{-}-\frac{1}{2}\{\sigma^{B}_{-}\sigma^{B}_{+},\rho^{AB}\})\nonumber\\
&&+\dfrac{\gamma_{2}(t)}{2}(\sigma^{B}_{-}\rho^{AB}\sigma^{B}_{+}-\frac{1}{2}\{\sigma^{B}_{+}\sigma^{B}_{-},
\rho^{AB}\}),
\end{eqnarray}
where $ \gamma_{1,2}$ shows time-dependent decay rates. The second and third terms describe heating and dissipation, respectively. 
\begin{figure}[t]
\includegraphics[scale=0.99]{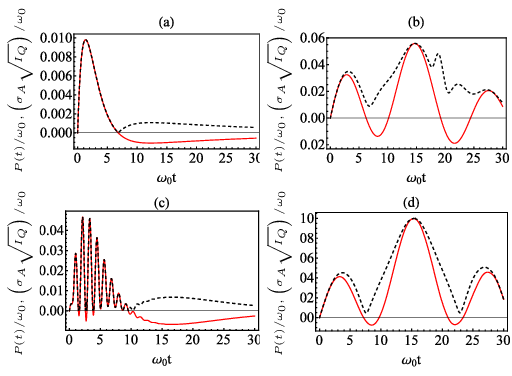}
\caption{(Color online). Plot of  $\sigma_{\mathscr{A}} \sqrt{I_{Q}}$ (dashed black line) and $\mathbb{P}$ (red solid line) (in units of $\omega_{0}$) as a function of $\omega_{0}t$ for $ T=0 $. Local Markovian dynamics for overdamped and underdamped regime is shown in (a) and (b), respectively. Also, local non-Markovian dynamics for overdamped and underdamped regime is illustrated in (c) and (d), respectively. 
The parameters are the same as Fig. \ref{fig3}.}
\label{fig4}
\end{figure}
 Suppose the spectral density of the environment is taken as
 \begin{eqnarray}\label{e44}
J(\omega)= \gamma_0 \lambda^2 / 2\pi[(\omega_0 - \Delta - \omega)^2 + \lambda^2],
\end{eqnarray}
in which $\gamma_{0}$ is an effective coupling constant related to the relaxation time of the battery system $\tau_{R}\approx1/\gamma_{0}$ and the width of the spectrum is presented by $\lambda$ connected to the reservoir correlation time $ \tau_{B}\approx1/\lambda$. Also, $\Delta =\omega_{0}-\nu_{c}$ is the detuning and $\nu_{c}$ is the central frequency of the thermal reservoir \cite{n1}. For example, an imperfect or leaky cavity is well approximated by such a spectrum.
By taking into account these considerations, the decay rates are given by $\gamma_{1}(t)/2=(N)f(t)$ and $\gamma_{2}(t)/2=(N+1)f(t)$, where $N= 1/ [\exp(\omega_{0}/k_{B}T)-1]$ represents the mean number of photons in the modes of the thermal reservoir at temperature $T$ and the function $f(t)$ depends on the form of the reservoir spectral density. Note that the heating rate vanishes at zero temperature, i.e., $\gamma_{1}(t)=0$, and the dissipation rate is determined by $\gamma_{2}(t)/2=f(t)$ \cite{n1}. The function $f(t)$  obtained in the exactly solvable form is given by \cite{n1}
\begin{eqnarray}\label{e42}
&&f(t)=-2 \Re\{\frac{\dot{C}(t)}{C(t)}\},\nonumber\\
&&C(t)= e^{-(\lambda-i\Delta)t/2}(\cosh(\frac{d t}{2})+\frac{\lambda-i\Delta}{d}\sinh(\frac{d t}{2}))C(0),\nonumber\\
\end{eqnarray}
with $d=\sqrt{(\lambda-i\Delta)^{2}-2\gamma_{0}\lambda}$. We can also define $R=\gamma_{0}/\lambda$ to distinguish the strong coupling regime from the weak coupling regime. It has been demonstrated that in the strong coupling regime, $R\gg 1$, the function $f(t)$ might take on negative values within certain time intervals, hence the dynamics of the qubit becomes nondivisible and non-Markovian \cite{n3, k}.
\begin{figure}[t] 
\includegraphics[scale=0.30]{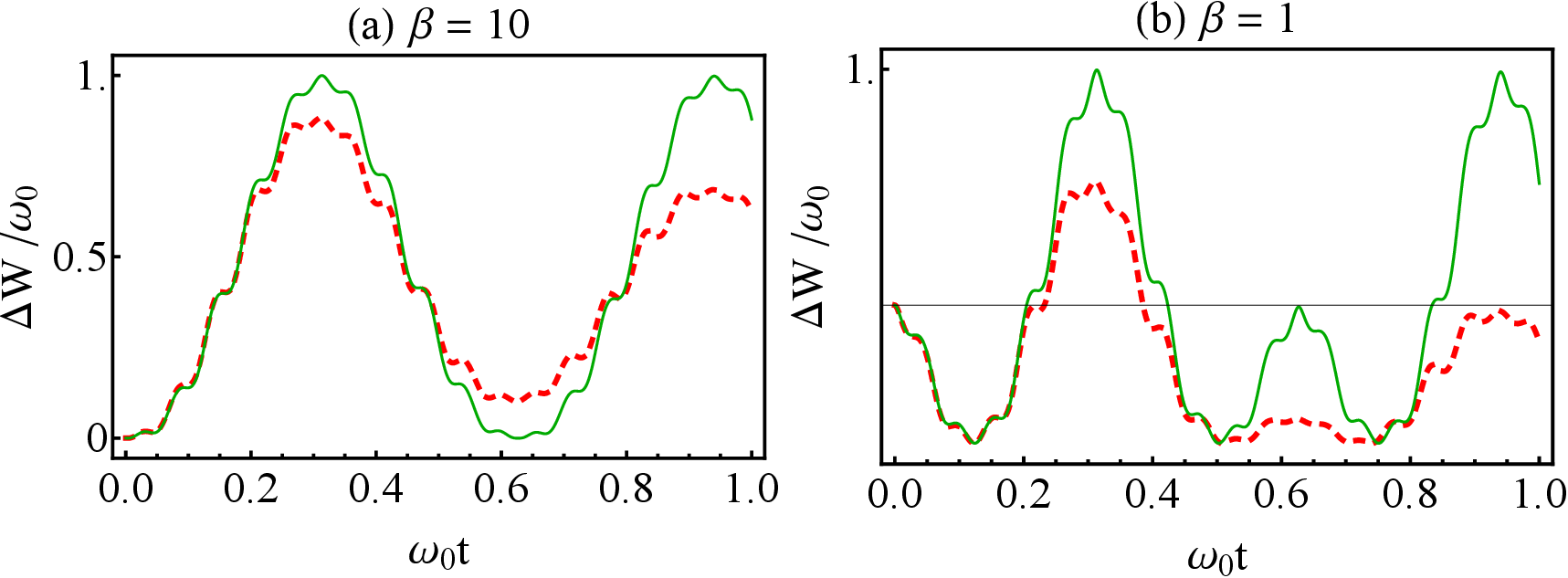}
\caption{Panels (a) and (b) show $\Delta W$ for underdamped regime at two inverse temperatures of \textcolor{black}{$\beta=1$} and \textcolor{black}{$\beta=10$}. In each panel, solid green line and dashed red line show non-Markovian and Markovian dynamics, respectively. Here, $N=5$, $\eta=10\omega_{0}$ and $\kappa=50\omega_{0}$. Other parameters are the same as those in Fig. \ref{fig3}.}
\label{fig5}
\end{figure}
In order to solve Eq. (\ref{e41}), we write $ \rho^{AB}$  in the matrix form
\begin{equation}\label{e43}
\rho_{AB}(t)=\begin{pmatrix}
    r_{11}(t) &  r_{12}(t) &  r_{13}(t)  & r_{14}(t)  \\
     r_{21}(t) &  r_{22}(t) &  r_{23}(t)  & r_{24}(t) \\
      r_{31}(t) &  r_{32}(t) &  r_{33}(t)  & r_{34}(t) \\
     r_{41}(t) &  r_{42}(t) &  r_{43}(t)  & r_{44}(t) \\
\end{pmatrix}.
\end{equation}
Substituting the above matrix into Eq. (\ref{e41}) gives a first-order system of ordinary differential equations in the sixteen unknown functions $r_{ij}(t)$, which has to be solved numerically under the initial conditions.

Next, change in the stored work, $\Delta W= W_{max} (t)-  W_{max} (0)$ in units of $\omega_{0}$, as a function $\omega_{0} t$ at $T=0$ is studied as presented in Fig. \ref{fig3}. The initial state is chosen as
 $|\varphi_{AB}(0)\rangle= |1\rangle\otimes|0\rangle$ implying that the $QB$ is empty. 
Solid blue line indicates overdamped regime and dashed red line presents underdamped regime that both of them are shown local Markovian dynamics. While dotted purple line remarks overdamped regime and solid green line displays underdamped regime which are considered for local non-Markovian dynamics. As can be seen,
the maximum value of stored work, i.e., $ \Delta W=\omega_{0}$, can be provided for underdamped and non-Markovian regime.  
\begin{figure}[t]  
\includegraphics[scale=0.28]{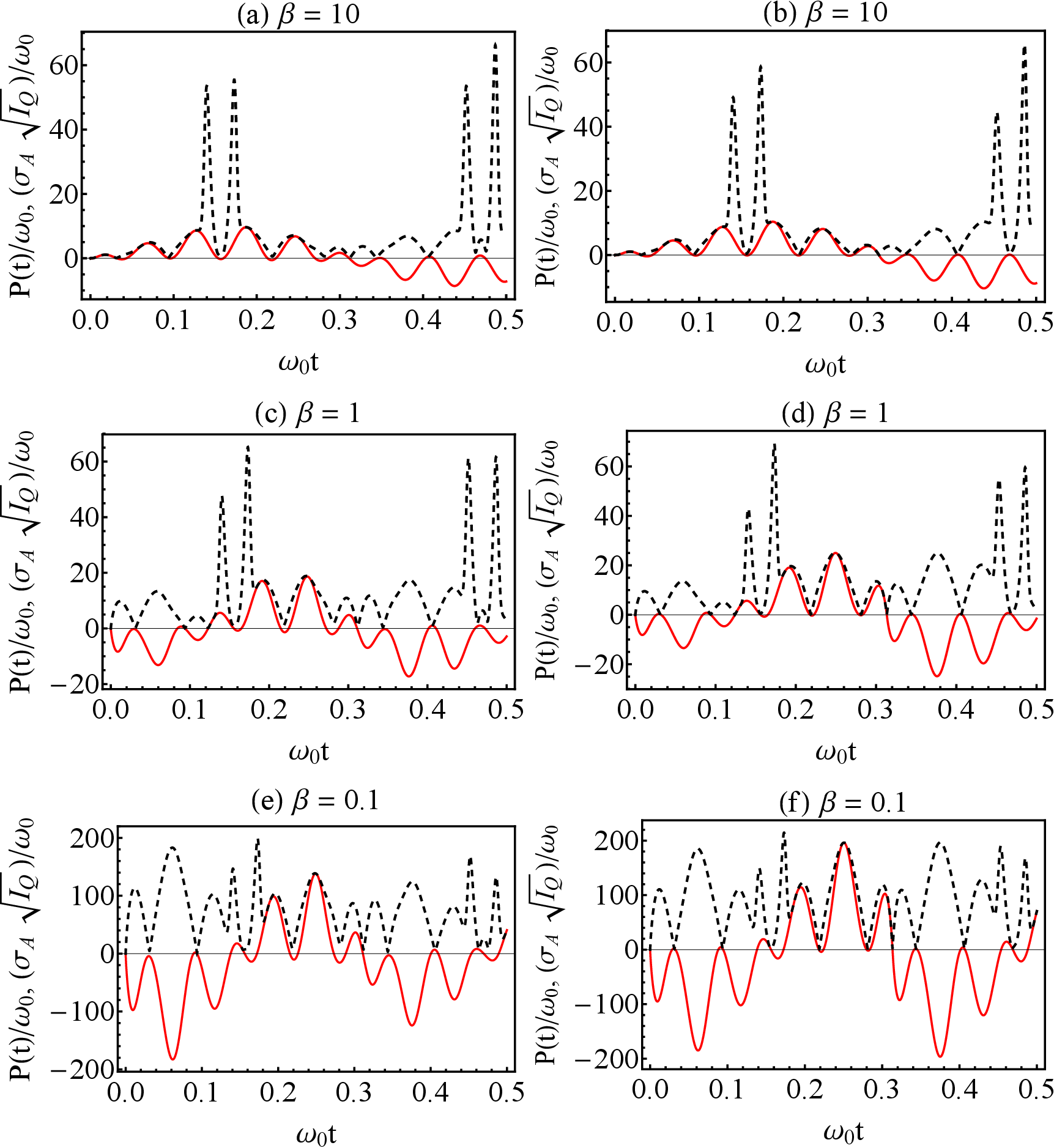}
\caption{In each panel, dashed black line illustrates $\sigma_{\mathscr{A}} \sqrt{I_{Q}}$ and solid red line shows $\mathbb{P}$, in units of $\omega_{0}$, as a function of $\omega_{0}t$. Panels (a), (c) and (e) for Markovian dynamics at the inverse temperatures of \textcolor{black}{$\beta=10$}, \textcolor{black}{$\beta=1$} and \textcolor{black}{$\beta=0.1$}, respectively. Panels (b), (d) and (f) illustrate non-Markovian dynamics at the inverse temperatures of \textcolor{black}{$\beta=10$}, \textcolor{black}{$\beta=1$} and \textcolor{black}{$\beta=0.1$}, respectively. Here, $N=5$, $\eta=10\omega_{0}$ and $\kappa=50\omega_{0}$. Other parameters are the same as those in Fig. \ref{fig3}.}
\label{fig6}
\end{figure}
In Fig. \ref{fig4}, charging power of the battery $\mathbb{P}$ and the upper bound $\sigma_{\mathscr{A}} \sqrt{I_{Q}}$ are plotted as a function $\omega_{0} t$ for $T=0$. Dashed black line presents $\sigma_{\mathscr{A}} \sqrt{I_{Q}}$ and red solid line shows $\mathbb{P}$.
Local Markovian dynamics for overdamped and underdamped regime is shown in Fig. \ref{fig4}(a) and 
Fig. \ref{fig4}(b) respectively. Also, local non-Markovian dynamics for overdamped and underdamped regime is illustrated in Fig. \ref{fig4} (c) and Fig. \ref{fig4}(d), respectively. 
Numerical results in panels (a)-(d) have been obtained by setting the parameters as Fig. \ref{fig3}. 

As the system undergoes a transition to the non-Markovian and underdamped regime (see Fig. \ref{fig4}(d)), the power
boosts significantly by up to two orders of magnitudes. It can be observed that the greatest value for charging power can be achieved for underdamped regime and non-Markovian dynamics at the time $\omega_{0} t=15$. Moreover, one can notice that at the same time, the power $\mathbb{P}$, reaches upper bound $\sigma_{\mathscr{A}} \sqrt{I_{Q}}$, implying that the bound is saturated (i.e., the power and upper bound curves closely meet.). It can be observed that the behavior of power and upper bound are qualitatively in agreement once the mentioned conditions are satisfied.

In order to investigate the role of temperature on the stored work, in Fig. \ref{fig5}, we have plotted $\Delta W$ at two two inverse temperatures \textcolor{black}{$\beta=1$} and \textcolor{black}{$\beta=10$}, respectively. \textcolor{black}{These selected temperatures are also considered in the literature \cite{beta1, beta2}. It is worth noting that at the considered temperatures, the quantum behavior of the system is still observed along with the temperature fluctuations.} Additionally, by taking into account the results from Figs. \ref{fig3} and \ref{fig4}, we have regarded only underdamped regime by choosing $\eta=10\omega_{0}$ and $\kappa=50\omega_{0}$. In each panel, dashed red line presents Markovian dynamics and solid green line shows non-Markovian dynamics. As can be observed, in non-Markovian dynamics at \textcolor{black}{$\beta=1$}, the stored work decreases then it becomes growing until reaches one, then the battery is fully charged at the time $\omega_{0} t=0.3$. Note that the amount of stored work varies between $ -1\leq \Delta W\leq 1$. Considering Eq. (\ref{e1}), it can be realized that the negative values are due to the temperature and the entropy effects. By increasing the temperature, the second term in free energy becomes a large negative term. Also, we should keep in mind that as the system dynamics evolve in contact with a thermal environment, the state of the battery tends to maximal mixed state (maximum value of Von Neumann entropy, 1). Therefore, most likely, the overall free energy takes a negative value, and the range of the energy variation strictly depends on temperature. \textcolor{black}{By Comparing the stored work in the non-Markovian and Markovian underdamped regimes, which are respectively shown by solid green and dotted red curves in Figs. \ref{fig5}(a) and (b) and Fig. \ref{fig3}, we can draw the following conclusions.
The curves in Figs. \ref{fig5}(a) and (b) become oscillating in the presence of finite temperature bath compared to their counterparts in Fig. \ref{fig3}. However, this behavior is less dominant in the non-Markovian regime. As can be observed from Figs. \ref{fig5}(a) and (b), in the non-Markovian underdamped regime (solid green line), in both panels, the battery is more charged compared to Markovian case and reaches the maximum value of 1. Also, by increasing temperature in the Markovian regime, the stored work further decreases and frequently fluctuates with negative amplitudes compared to its behavior in the non-Markovian regime. From the battery stabilization viewpoint, non-Markovianity makes the battery stored work more robust against temperature.}

\begin{figure}[t]  
\includegraphics[scale=0.28]{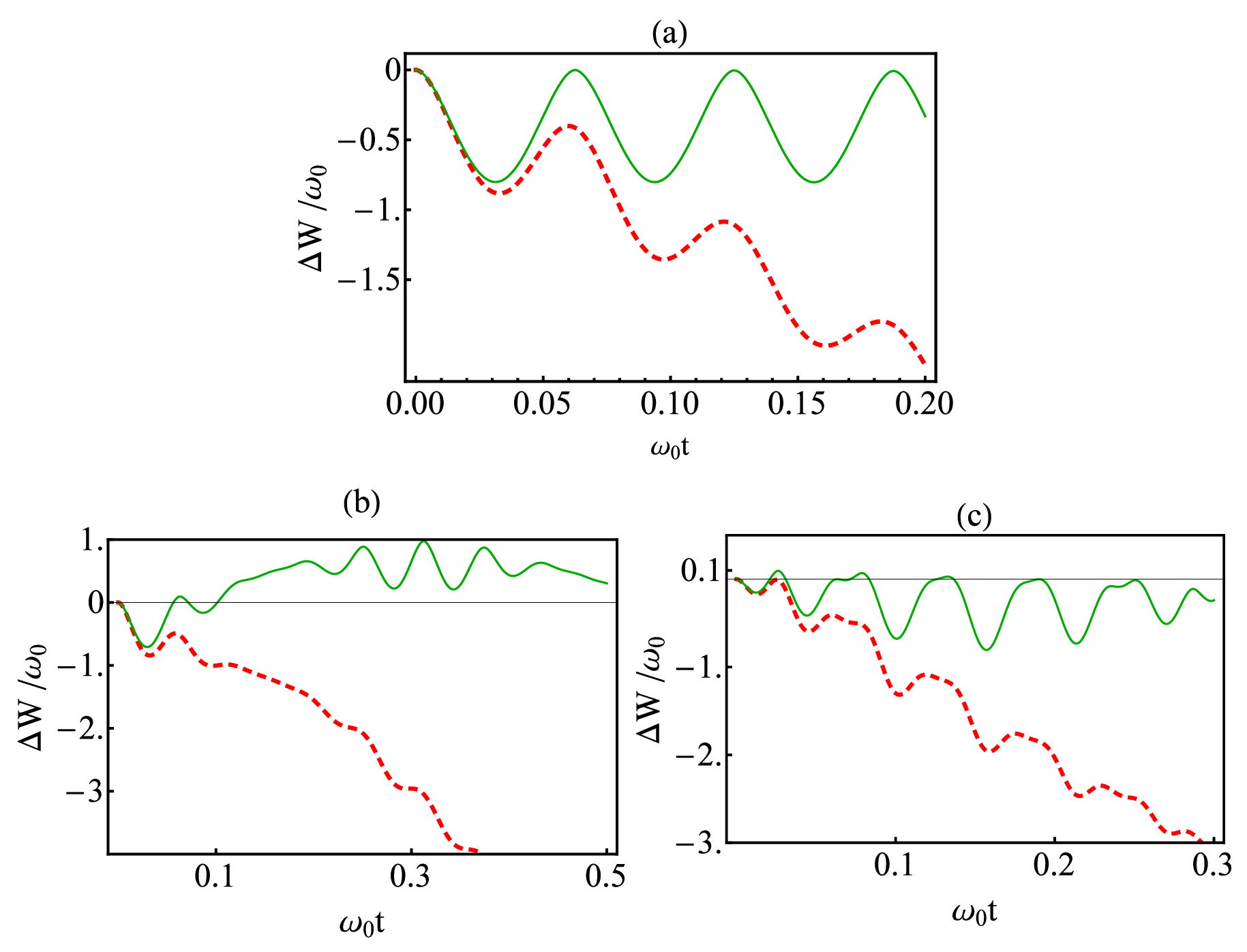}
\caption{  $\Delta W$ for different initial states. Solid green line and dashed red line represent non-Markovian and Markovian dynamics respectively. Panel (a) $|\varphi_{AB}(0)\rangle=\frac{1}{\sqrt{2}}(|0\rangle+| 1\rangle)\otimes \frac{1}{\sqrt{2}}(|0\rangle+| 1\rangle) $. Panel (b) $|\varphi_{AB}(0)\rangle=\frac{1}{\sqrt{2}}(|0\rangle+| 1\rangle)\otimes| 0\rangle$. Panel (c) $|\varphi_{AB}(0)\rangle=| 1\rangle\otimes \frac{1}{\sqrt{2}}(|0\rangle+| 1\rangle) $. Parameters are the same as those in Fig. \ref{fig5}} 
\label{fig7}
\end{figure}

The effect of temperature on charging power is illustrated in Fig. \ref{fig6}, where in each panel, the power and its corresponding upper bound are shown by red solid line and dashed black line, respectively. By comparing the results obtained from three bath temperatures, we can deduce the following observations. The magnitude of charging power, is higher in the non-Markovian regime compared to that in the Markovian regime, especially at high temperatures. \textcolor{black}{By increasing temperature, the behavior of charging power becomes more oscillating with a larger fluctuation amplitude. However, the upper bounds offer tighter estimations of the charging power in the non-Markovian regime compared to their Markovian counterparts at all considered temperatures. This observation is more evident at higher temperatures.}

Finally, the effect of initial coherence on the stored work is investigated. For this, $\Delta W$ is depicted for different  initial states in Fig. \ref{fig7}. We consider the initial state as $|\varphi_{AB}(0)\rangle=\frac{1}{\sqrt{2}}(|0\rangle+| 1\rangle)\otimes \frac{1}{\sqrt{2}}(|0\rangle+| 1\rangle) $ in Fig.\ref{fig7}(a), that there is initial coherence in both of the charger and the battery. Fig. \ref{fig7}(a) shows the value of the stored work is always negative and its maximum value is zero, accordingly, the battery can not be charged. We assume $|\varphi_{AB}(0)\rangle=\frac{1}{\sqrt{2}}(|0\rangle+| 1\rangle)\otimes| 0\rangle$ and $|\varphi_{AB}(0)\rangle=| 1\rangle\otimes \frac{1}{\sqrt{2}}(|0\rangle+| 1\rangle) $ as initial states in Fig.\ref{fig7}(b) and Fig. \ref{fig7}(c), respectively. By comparison panel (b) and (c), we see the battery can be charged completely, i.e., $\Delta W= \omega_{0}$, in the absence of initial coherence in the battery as well as the existence of initial coherence in the charger has no  \textcolor{black}{constructive} effect on the amount of stored work. \textcolor{black}{In the Markovian regime, this effect is observed regardless of selecting
an initial coherence in the subsystems of battery or charger, as shown in Figs. \ref{fig7} (a) and (b). The effect is also seen in the non-Markovian regime when the initial coherence exists in the battery subsystem (see Fig. \ref{fig7} (c)).}

Remark. Some considerations to apply Eqs. \ref{e3} and \ref{e6} in the present example are as follows. In  Sec. II below the Eq. \ref{e1}, $\rho_{eq}$ is defined by known canonical equilibrium or Gibbs state  $\rho_{eq}\equiv\tau_{\beta}=\frac{1}{Z} \exp (-\beta H_{B})$. By this definition, an important assumption from the thermodynamic point of view is made. Accordingly, the ratio of decay rates satisfies the property of local detailed balance, which follows from the Kubo-Martin-Schwinger (KMS) relation of the reservoir correlation function. It implies sufficient condition to make the Gibbs state become the stationary solution of the equation of system evolution. This condition is fulfilled in weak coupling approximation. However, it is shown \cite{strong1} that the 'energy conservation' hypothesis of thermal operations is achieved by making some proper assumptions on total dynamics. For example, one way to achieve this is to use a time-dependent interaction that is switched on and off. As a consequence, the Gibbs state of the system can be a fixed point of the dynamical map even beyond the weak-coupling limitations. Getting back to our present example, non-Markovianity arises from Existing narrow structures appearing in the spectrum of the reservoir. Depending on the value of parameters taken for the Lorentzian spectral density function, both Markovian and non-Markovian regimes could govern the dynamics. However, in the strong coupling regime, the Gibbs state $\tau_{\beta}$ may not exact stationary state with respect to the bare Hamiltonian of the system. In fact, in a study performed by Strasberg et al. \cite{strong2}, it has been shown that in the strong coupling regime, for short times the ratio of rates does not fulfill local detailed balance. Hence, the instantaneous fixed point may deviate from the Gibbs state. Instead, the generalized Gibbs state can be obtained using mean-force Hamiltonian. Also, it has been stated that similar situation may occur for Markovian evolution with positive time-dependent decay rates even if the underlying Hamiltonian is time-independent. However, it has been demonstrated that rates become stationary for long times, and their ratio fulfills local detailed balance. It implies that the steady state is a Gibbs state, and hence, the system properly thermalizes. In conclusion, in the case of this example, Gibbs state is not the exact instantaneous fixed point, but surely it is the equilibrium state which the system tends to it after some while.

\section{Conclusion}
In summary, we have studied bounds on the charging power of quantum batteries via an open system approach.
Having introduced an activity operator, which quantifies how far the state of the system distances from equilibrium, a tight upper bound on charging power has been proposed in terms of quantum Fisher information and the variance of the activity operator. The former describes the speed of evolution, and the latter may be interpreted as a generalized thermodynamic force. In addition, we have obtained a thermodynamic interpretation of the power in terms of dissipative work and the rate of irreversible entropy. In the following, by applying the notion of the extended QFI, an upper bound on the charging power has been suggested. The advantage of this task is that it allows us to divide the speed of evolution into the classical and quantum parts. Hence, a description based on more physically meaningful contributions has been proposed.
To evaluate the introduced bound we have investigated two examples. We have first considered the Heisenberg XX spin chain to illustrate the bound is tight. Taking into consideration the environmental effects, in the second example, we have demonstrated that the battery can be fully charged in the non-Markovian dynamics and underdamped regime and its power is also greater than the Markovian case. The results show that, under the above conditions, the behavior of power and upper bound over time are in close agreement.
Also, it has been indicated that the charging power increases by increasing the temperature. Moreover, our results show that, at the very least, the existence of initial coherence does not show any remarkably constructive effect on the amount of the stored work. These findings along with the implementation of reservoir engineering techniques may pave a way to develop batteries with more powerful charging and also more stable and controllable performance.

\section*{ACKNOWLEDGMENTS}
This work has been supported by the University of Kurdistan. Authors thank Vice Chancellorship of Research and Technology,  University of Kurdistan.
%%%%%%%%
\renewcommand{\appendixname}
{APPENDIX}
\appendix

\section{CORRESPONDENCE BETWEEN TWO FORMULATIONS OF THE ACTIVITY OPERATOR}
\label{appendix A}

In order to obtain Eqs. 22 to 25, we have to decompose the activity operator $\mathscr{A}=\beta^{-1}log \left ( \rho/\tau_{\beta}\right )$ into these three terms $\mathscr{A}= H_{B}+\beta^{-1} \log \rho+F_{\tau_{\beta}}$ . Here, some straightforward algebras can show the correspondence between the two formulas.

\begin{align}\label{c1}
\frac{1}{\beta} \log \frac{\rho}{\tau_{\beta}}&=\frac{1}{\beta} \log \rho-\frac{1}{\beta} \log \tau_{\beta}\\ \notag
&=\frac{1}{\beta} \log \rho-\frac{1}{\beta} \log _{e} \frac{e^{-\beta H_{\text {B}}}}{\mathbb{Z}}\\ \notag
&=\frac{1}{\beta} \log \rho+H_{\text {B}}-\frac{1}{\beta}\log _{e} \mathbb{Z}\\ \notag
&=\frac{1}{\beta} \log \rho+H_{\text {B}}+F_{\tau_{\beta}},
\end{align}

in which, $\tau_{\beta}=\frac{e^{-\beta H_{B}}}{\mathbb{Z}}$ is the thermal state of the battery, and $F_{\tau_{\beta}}$ is the equilibrium free energy. The above calculation may lead to some ambiguity raising from adding up non-operator and operator terms. In order to remove this inconsistency we may deal with the term $\frac{1}{\beta} \log _{e} \frac{e^{-\beta \hat{H}_{B}}}{\mathbb{Z}}$ as follows.

\begin{align}\label{c2}
\frac{1}{\beta} \log _{e} \frac{e^{-\beta \hat{H}_{B}}}{\mathbb{Z}}&=\frac{1}{\beta} \log _{e} \frac{1}{\mathbb{Z}} \hat{\mathbf{1}} e^{-\beta \hat{H}_{B}}\\ \notag
&=\frac{1}{\beta} \log _{e} \frac{1}{\mathbb{Z}} \hat{\mathbf{1}}+\frac{1}{\beta}\log _{e} e^{-\beta \hat{H}_{B}} \\ \notag
&=\left(\frac{1}{\beta} \log _{e} \frac{1}{\mathbb{Z}}\right) \hat{\mathbf{1}}- \hat{H}_{B}=-F_{\tau_{\beta}}\hat{\mathbf{1}}- \hat{H}_{B}.
\end{align}

\section{COMPARISON BETWEEN TWO BOUNDS}
\label{appendix B}

In order to evaluate the tightness of the bound presented in Sec. II, $\sqrt{I_{Q}} \sigma_{\mathscr{A}}$, we compare it with a previously suggested bound in Ref. \cite{op1}. In the following, a proof allowing the comparison is presented. First, let us briefly introduce the bound presented by Eq(12) of Ref. \cite{op1},
\begin{equation}\label{g1}
|\mathbf{P}| \leq \sigma_{\mathcal{F}} \sigma_{\mathcal{V}},
\end{equation}\
where $\sigma_{\mathcal{F}}$ and $\sigma_{\mathcal{V}}$ are respectively standard deviations of the defined free energy operator $\mathcal{F}=H_{\mathcal{W}}+\frac{1}{\beta}\log\rho_{\mathcal{W}}$ and of the battery interaction Hamiltonian. The operator $\mathcal{F}$ and our activity operator $\mathscr{A}$ are not exactly equivalent. However, a straightforward calculation shows that their standard deviations are exactly the same, $\sigma_{\mathscr{A}}=\sigma_{\mathcal{F}}$. So, evaluating the tightness between two bounds reduces to compare two terms of $\sigma_{\mathcal{V}}$ and $\sqrt{I_{Q}}$. We proceed as follows
, noting that that in our work interaction Hamiltonian is denoted by $H_{int}$, but for the ease of comparison, we choose the same notation, $V$, as Ref. \cite{op1}.
According to Eqs. (8) and (9), the Fisher information can be written as
\begin{equation}\label{g2}
I_{Q}=\operatorname{Tr}\left(\rho L^{2}\right)=\operatorname{Tr}(\dot{\rho} L).
\end{equation}\
The evolution of the state of battery in the interaction picture is written as $\dot{\rho}=-\frac{i}{\hbar}~Tr[V, \rho_{tot}]$, in which $\rho_{tot}$ denotes the total state. Replacing in the above equation yields

\begin{align}\label{g3}
I_{Q}= &-i ~Tr([V,\rho_{tot}]L)= -i ~Tr([V-\langle V\rangle,\rho_{tot}]L) \\ \notag
& =-i~ Tr([\delta V,\rho_{tot}]L),
\end{align}
where, in the second line, we define $\delta V=V-\langle V\rangle$. Considering triangle and Cauchy-Schwarz inequalities, the following equations can be obtained

\begin{align}\label{g4}
I_{Q}^{2} & \leq\left|\operatorname{Tr}\left(\delta V, \rho_{t o t}\right) L\right|^{2}=\left|\operatorname{Tr}\left(\delta V \rho_{t o t} L-\rho \delta V L\right)\right|^{2} \\ \notag
& \leq\left|\operatorname{Tr}\left(\delta V \rho_{t o t} L+\rho \delta V L\right)\right|^{2} \\ \notag
& \leq\left(\left|\operatorname{Tr}\left(\delta V \rho_{t o t} L\right)\right|+\left|\operatorname{Tr}\left(\rho_{t o t} \delta V L\right)\right|\right)^{2} \\ \notag
&=4\left|\operatorname{Tr}\left(\delta V \rho_{t o t} L\right)\right|^{2},
\end{align}

where the last step results from the cyclic property of trace. By utilizing the Cauchy-Schwarz inequality

\begin{align}\label{g5}
I_{Q}^{2} & \leq 4\left|\operatorname{Tr}\left(\rho_{\text {tot}} \delta V L\right)\right|^{2} \\ \notag
& \leq 4 \operatorname{Tr}\left(\rho_{\text {tot}}(\delta V)^{2}\right) \operatorname{Tr}\left(\rho L^{2}\right)=4 \sigma_{V}^{2} I_{Q},
\end{align}

Note that the trace in the term $\operatorname{Tr}\left(\rho L^{2}\right)$ reduces to partial trace over battery subsystem. Therefore, we can conclude that

\begin{equation}\label{g6}
\sqrt{I_{Q}}\leq 2 \sigma_{H_{int}}.
\end{equation}\

The above inequality demonstrates the comparison between the tightness of two presented bounds,

\begin{equation}\label{g7}
\sigma_{\mathbb{F}} ~\sqrt{I_{Q}}\leq 2 \sigma_{\mathbb{F}} ~\sigma_{H_{int}}.
\end{equation}\

%%%%%%%%%%%%%%%%%%%%%%%%%%%%%%%%%%%%%%%%%%%%%%%%%%%%%%

\end{document}